# A Review of Multi-material and Composite Parts Production by Modified Additive Manufacturing Methods

M. Toursangsaraki [a]

[a] State Key Laboratory of Mechanical System and Vibration, School of Mechanical Engineering, Shanghai Jiao Tong University, Shanghai 200240, China

## Abstract

Aside from the capability of additive manufacturing (AM) methods in fabricating components with complex geometries, two crucial potentials of this manufacturing process that are worth mentioning are its flexibility in being combined with other production methods as well as use of a variety of materials in a single production platform to make multi-material and composite products. Implementation of multiple materials in integrated structures has been shown to improve the functionality, weight reduction and, by merging the assembly and production into one stage, modify the manufacturing processes. Different approaches towards modification of AM processes aimed to reach multi-material or composite parts are being reviewed in this paper.

# Contents

# 1. Introduction

A unique feature of Additive Manufacturing (AM) technology is production capability of multi-material parts. In this approach, multiple types of materials can be used for fabrication of a single part. Components with specially tailored functionally graded, heterogeneous or porous structures and composite materials have been some of the achievements of this method [1-3]. A wide range of materials such as metals, plastics, and ceramics has been used in various AM methods to obtain multi-material products in order to match the current requirements of the industry which wouldn't be gained otherwise. All the AM techniques have the potential of being applied to multiple material manufacturing in nature. Moreover, many studies have been performed to investigate the possibility of applying multi-material production for different AM methods [4, 5]. The process of making composite materials by AM can either be performed during the material deposition process or by a hybrid process in which the combination of different materials can be performed before or after AM as a previous or subsequent stage of production of a component. Composite processes can be implemented to produce heterogeneous scaffolds and functionally graded materials (FGM). Production of tailor-made gradient multi-phase or porous materials is one of the features of AM processes. As a result, different properties can be achieved within one single integrated part. Furthermore, in making a component with composite materials, the required properties of included materials can be combined while compensating for some of their restrictions.

# 2. Multi-material and composite additive manufacturing methods

Different modifications of AM and combination of them with other manufacturing methods aimed to generate multi-material or composite products have been listed below. The subjects are mostly categorized by the process adaptations to implement multi-materials, different materials being used in these processes, and hybrids of AM and other manufacturing methods.

## 2.1. Stereolithography methods

Stereolithography (SLA) has been developed based on photopolymerization phenomena and mostly involves implementing a light source to bond photo-curable resins mixed with other materials to manufacture solid composite parts [6, 7]. Multi material SLA processes have been performed by successively applying and washing off different kinds of photopolymerizable resins in single or multiple vat setups (Fig 1.a) to fabricate each piece of the component with the specific desired materials. The functionality of these approaches has been spread in several fields from fabricating electronic parts to biomedical implants [1, 8-10].

The liquid precursor infiltration method has been typically performed for production of ceramics and their composites. In this method, a porous component is immersed in a liquid infiltration material. As a result of infiltration of the precursor into the pores of the ceramic component, a wide variety of microstructures such as gradient, partially or fully dense materials as well as an increase in density and mechanical properties of the powder compact can be achieved [11, 12]. In SLA of ceramic materials, a ceramic suspension including photocurable liquid resin is used to produce the green part in the desired shape. Drying and other debinding processes are then implemented to achieve a component with high density and minimal defects [13], Liu et al. [14] performed an infiltration process after SLA of alumina ceramics by immersing the printed products into liquids of different modifying components to make a multi-phase composite by filling the interconnected porosities with infiltrated materials. A photo-initiator material was added to alumina suspension to make it UV-curable and suitable for SLA process. Debinding and subsequent infiltration were followed by a precipitation process (Fig 1.b). As a result of the infiltration process, an increase in the hardness of the part but a decrease in fracture toughness was reported.

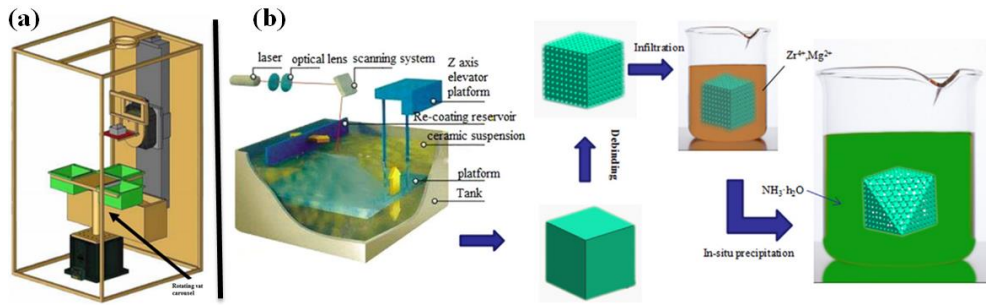

**Fig 1**. (a) Multiple vat setup for stereolithography of multi-material parts [10], (b) SLA followed by infiltration and precipitation process [14].

## 2.2. Binder jetting methods

Binder jet printing deposits binder materials on powder bed to selectively join powder materials layer by layer to construct three-dimensional parts [15, 16]. In binder jetting, additional extractable powder materials can be engineered in the binding process to reach a desired percentage of materials in different layers of parts. Then by the use of extraction procedures such as solvent materials, additional materials can be removed to obtain a desired porous or functionally graded material (FGM) product [17].

Applying infiltration processes to ceramic materials produced by AM has been utilized to create multiphase composites. As for the binder jet process, the process uses the nature of porous products of binder jetting. While the binder material is cured to achieve a solid part, they burn off and leave porosities in the final component. The main purpose of infiltration procedure is to fill the porosities with other functional materials to achieve a fully dense part [18, 19]. Ceramic-metallic composites can be produced by binder jetting the ceramic powder materials, curing and sintering them in specific conditions in order to achieve a solid homogeneous structure and immersing them in a molten metal bath to fill the porosities with metal. The sintering temperature of the printed ceramic part can affect the density and volume fraction of metal phase and consequently microstructure and mechanical properties of the final part [20]. The submerging time of sintered ceramic part in the molten metal is another discussable parameter that can change the properties of the products as the more sintered ceramic parts are held in molten metal, apart from infiltration process, more metallic material permeates into the ceramic particles and the properties of the products will have higher similarity to the pure metallic parts made of the molten material [21]. This methodology has also been used for fabrication of metallic composites. In one study, it has been illustrated that the layer thickness of printed part has a crucial effect on different properties of the final part as it changes the microstructure and chemical composition of the product [22].

## 2.3. Extrusion-based printing methods

Extrusion-based AM uses an extrusion nozzle to deposit materials on a substrate to print the desired component layer by layer. Low-temperature deposition manufacturing (LDM) is an extrusion-base AM method which is capable of manufacturing multi-material components with custom-built porosities (Fig 2). Because LDM process is operated at low temperature, the bioactivity and biocompatibility of biomaterials can be preserved. This process is capable of manufacturing scaffolds with functionally graded or composite materials. Biomaterial scaffolds made of synthetic and natural polymers in tissue engineering applications such as scaffolds mimicking bone and cartilage structure or nerve and vascular tissues are the main focus of this fabrication technique [23, 24]. The procedure typically includes depositing the mixture of materials on an ultra-low temperature platform by a sterilized syringe followed by a freeze-drying process to remove the solvent material and diminish micropores generated by phase separation. LDM made scaffolds with core-shell composite framework have been created to improve their mechanical and physiochemical properties. An inner and outer feedstock tube and nozzle head were assembled together to extrude the core and the sheath material simultaneously [25]. A multi-nozzle LDM system using disposable syringes has been applied in order to fabricate scaffolds with gradient biomaterials and functions for tissue engineering [26, 27].

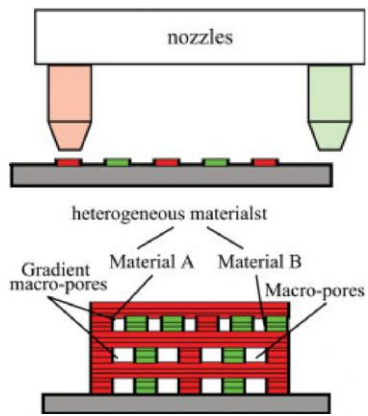

**Fig 2**. Schematic of multiple nozzle extrusion based AM of gradient materials [26].

Fused deposition modeling (FDM) uses filaments containing thermoplastic polymers which are melted and extruded through a nozzle on the desired substrate layer by layer. One of the polymers being fabricated by this method is polyvinylidene fluoride (PVDF), which application has been expanded recently in energy harvesting systems [28]. In FDM of ceramic AM, a composite of polymer and ceramic can be used as the feedstock material to improve the physical and mechanical properties of the final component [29, 30]. Kalita et al. [31] implemented FDM to fabricate composite scaffolds of PP and TCP with tailored interconnected porosities. Different scaffold architectures and the compatibility of the scaffold materials for vitro cell culture were evaluated and the results indicated the ability of the process to manufacture biomaterial components. The integrity and uniformity of the composite filament for the extrusion process is a pivotal factor in accomplishing the desired properties of the products. Other AM-related parameters such as raster gap and width as well as slice thickness are effective in the scaffold structure and, hence, mechanical properties and dispersion of porosities.

The use of sacrificial polymeric mixture or UV curable resins combined with ceramic particles is common in many AM methods of ceramic components production. The additive mixtures roles as a binder in the AM process to produce a green product and gets removed from the part by a subsequent debinding process. However, every single of these approaches usually suffers from its own certain obstacles such as excessive material consumption, low density after sintering, need for additional steps, lack of functionality and limitations in the produced component dimensions. An addition of photopolymerizable dispersion to the raw material of the extrusion-based AM has been tried to take advantage of the steadiness of the green product with UV-cured resin and economical syringe-base AM process. The UV-light irradiation is implemented during the printed process and UV-resin is removed by a typical sintering process. However, deficiencies such as partial polymerization of the layers and cracking of the part due to high shrinkage ratio and during sintering are yet to be dealt with [32].

## 2.4. Material jetting printing methods

The principals of ink-jet printing (IJP) typically include deposition of a jet stream of droplets or particles of materials, so the materials can fuse and bond together [33, 34]. Different actuating systems for ejection of droplets have been investigated including thermal bubble method, piezoelectric nozzle head, pneumatic diaphragm actuator, and the use of the electrical field to generate ink flow [35, 36]. The application of multiple nozzles or mixing raw materials for printing in the print head, allow the production of multi-material, multiphase composites and FG materials [37-39]. A combination of tape casting technique and IJP has been applied to fabricate electrolyte and electrode of micro-batteries [40, 41]. Flexible supercapacitors have been fabricated by IJP of graphene-based materials on metal films [42]. Ink-jetted dielectrics have also been performed to print insulators for packaging of electronic embedding devices [43] and securing metallic materials crossovers [44]. IJP has been utilized to deposit photoactive materials layer for solar cells [45] and multiple materials of photodetectors, both attempts in a combination of applications such as spin coating deposition method [46]. A combination of reverse offset printing, IJP and bar coating was performed for fabrication of flexible phototransistor. In this method, IJP is responsible for polymeric active semiconductor and conductive polymer for gate electrodes [47].

Multi-jet modeling (MJM), also known as material jetting system or poly-jet printing of materials, uses multiple jet nozzles to deposit photopolymers for the part structures, which are immediately cured after the deposition, and gel-like wax materials for the sacrificial support structures [48, 49]. A schematic depiction of the process is shown in Fig 3. This AM technique has the capability to fabricate components with higher resolutions and geometrical complexity such as microfluidic devices with narrow gaps and high-aspect ratios [50-52].

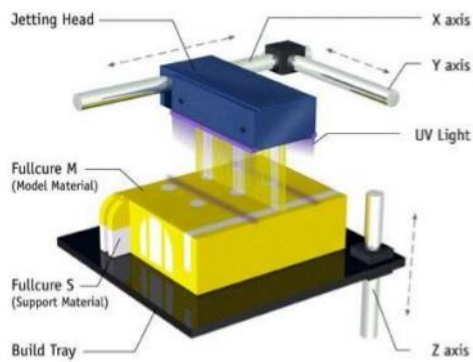

**Fig 3**. Schematic of poly-jet 3D printing [53]

Implementing electrospinning technique in syringe-base AM production has been performed in order to reach micro/nano-scale fiber diameters. Utilizing electrospinning in biomaterial AM fabrication methods can lead to better mechanical and biological properties and promote cell proliferation within the tissue engineering scaffolds [54-57]. Some of the limitations of electrospinning methods such as buckling and coiling of the jet stream have been overcome by performing hybrid process with melt-electrospinning [57, 58]. A Combination of syringe-based AM and wet-spinning process, which includes extruding a polymer mixture filament into a precipitating bath to solidify and produce continuous composite fibers, has been performed to produce microfibers for microporous composite scaffolds [59-61]. Because there is no need for high temperature, high voltage or toxic solvents, wet spinning allows for the production of fibers including biomolecules, which makes it suitable for bioprinting. Fig 4 illustrates the setup of electro-spinning and wet-spinning processes.

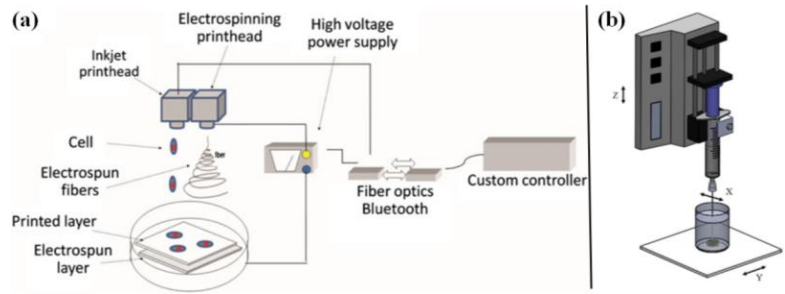

**Fig 4**. (a) Hybrid of inkjet printing and electrospinning [54], (b) setup of wet-spinning method [59]

## 2.5. Metallic alloys and functionally graded materials (FGM) AM techniques

### 2.5.1. Powder bed fusion methods

The use of laser or electron beam energy source to selectively melt or sinter powder materials to join them together layer by layer is another AM method called powder bed fusion technique mainly including Selective Laser Melting (SLM), Selective Laser Sintering (SLS), and Electron Beam Melting (EBM) [62-66]. This technique is capable of processing metals, ceramic and polymers [67-69]. Fabrication of multi-material products and functionally graded materials, which mechanical properties varies continuously through the thickness of the fabricated sample based on a known exponential function, is usually performed by using multiple chambers with different powder materials to be deposited on different layers in order to make the desired FGM component [70-74].

### 2.5.2. Directed energy deposition (DED) methods

Directed energy deposition (DED) uses a focused energy source to melt and deposit feedstock. The heat source used in DED process is usually provided by laser, electron beam or electric arc. Moreover, powdered or wired materials are typically used as feedstock materials. This process has been primarily implemented for the metallic parts production [75-77].

Production of bimetal of steel and bronze has been performed by use of Gas Metal Arc Welding (GMAW) to melt and deposit steel and bronze wired materials [77]. For DED of metallic alloys or composites, pre-alloyed feedstock materials are typically used. However, the desired compositions could be achieved by mixing the elemental powders with adjustable fractions during the process. The elemental powder materials can be pre-mixed and fed through a single powder feeder [78] or be fed separately from multiple coaxial power feeders with independently controllable feed rates that lead the various powder flows to converge to the melting point and be deposited on the substrate [79, 80]. The latter is suitable for the production of functionally graded composites, which is used in spacecraft multilayered structures [79, 81-84]. Fig 5 illustrate schematics of multi-chamber SLS and multi-nozzle DED.

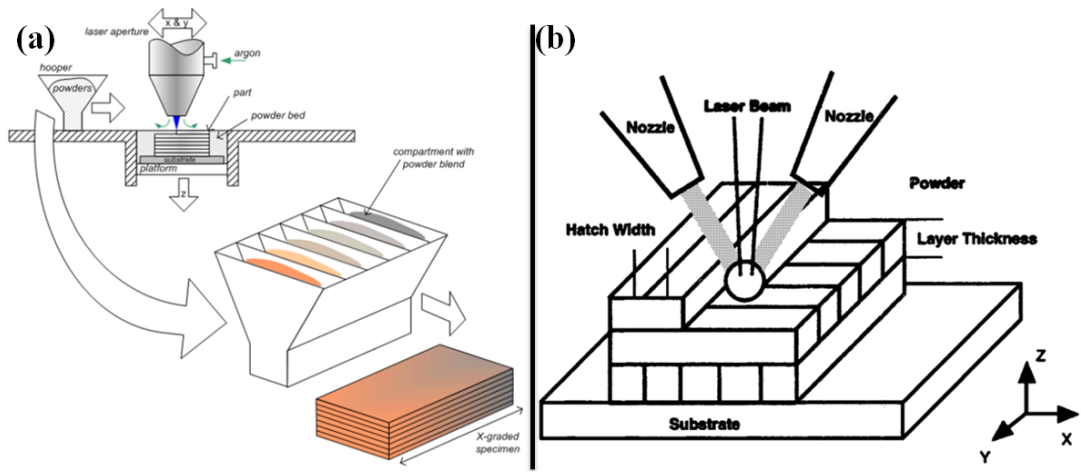

**Fig 5**. (a) Configuration of printing graded materials by SLS [85], (b) multiple nozzle DED [86]

### 2.5.3. Laminating metallic parts

In Laminated Object Manufacturing (LOM), three-dimensional components are fabricated by a combination of forming and lamination of sheets layer by layer. In LOM of metallic parts, laser or mechanical cutters are usually used for forming sheets and bonding process of different layers of sheet materials is achieved by chemical adhesives, thermal bonding, or brazing and welding approaches. Then a finish machining process is applied to achieve the desired geometrical accuracy and surface finish [87, 88].

Ultrasonic welding (UC), a solid-state metal seam welding method, has been widely utilized for sheet metal lamination methods. In this process, an energy produced by an ultrasonic transducer is transferred to the thin metal foils, which are deposited layer by layer, by a rotating sonotrode. The sonotrode travels on the upper surface of the component and oscillates in the vertical direction with high frequency and low amplitude causing diffusion of metal atoms at the sheets' interface and bond them together [89, 90]. Due to the low-temperature operation of UC process, production of metallic multi-material FGM [91-93], metal matrix composites [94-97], and structures with embedded electronics [98] are possible.

Friction welding and friction stir welding are two kinds of solid-state welding methods which have been widely used for welding of similar and dissimilar metallic materials [99-101]. These processes allow for layer-wise lamination of metallic parts which is then followed by a machining process to shape the layers and produce the desired three-dimensional component [102, 103].

## 3. Hybrid applications

## 3.1. Bioprinting methods

Bioprinting technique is usually referred to as a hybrid process that allows printing of tailored structures including multiple cells and biomaterials in a single part [104, 105]. IJP, extrusion-based printing and laser-aided printing are the most common methods used for bioprinting applications. Bioprinting allows for the production of a wide variety of practical biomedical tissues with different shapes and material compositions [106-108]. Fig 6 demonstrates an integrated bioprinting of cells, biomaterials and biological molecules applied for fabrication of tissues and organs.

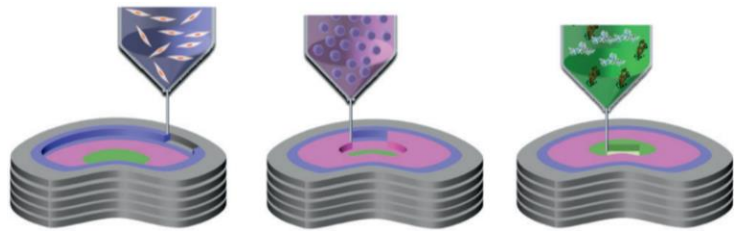

**Fig 6**. A schematic of bioprinting of tissues and organs [104].

## 3.2. Shape deposition methods

As mentioned before, relatively low dimensional accuracy and poor surface finish quality are the two most critical limitations of rapid prototyping methods. The hybrid of subtractive and additive manufacturing has been accomplished to incorporate the design freedom with precision within a fabrication process. Shape deposition manufacturing (SMD) uses a sequence of deposition and subtractive shaping of materials which allows for production and assembly of components at the same time as well as embedding internal parts inside every each component during the process [109, 110]. Having the abilities of the production of multi-material structures with varying material properties and embedding prefabricated components such as electronic parts like sensors and actuators within the components, SDM is capable of processing biomimetic mechanisms such as compliant joints and rigid parts of adaptive and robust robotic limbs [111-114]. An example of polymer-based SDM to produce a heterogeneous monolithic robotic grasper structure as well as a comparison between the SDM made and conventional robotic limb mechanism is shown in Fig 7.

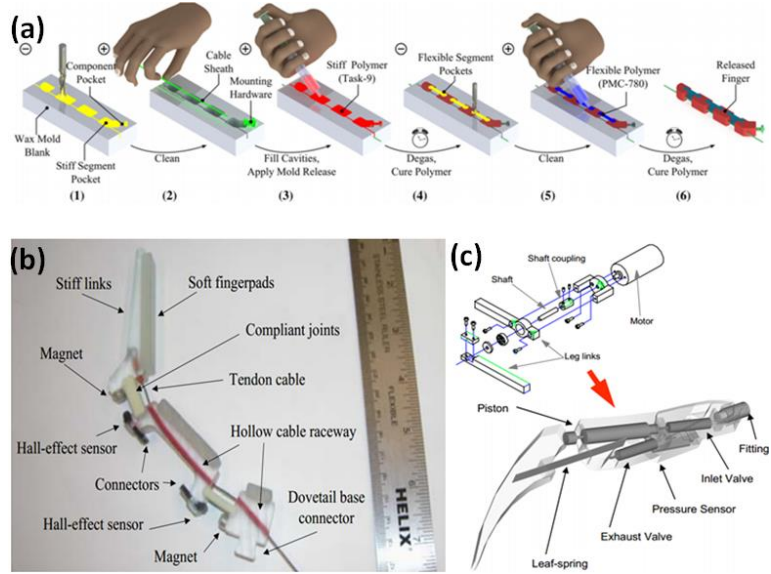

**Fig 7**. An example of polymer based SDM for production of a heterogeneous monolithic robotic grasper structure (a) and a comparison between the SDM made (b) and conventional robotic limb mechanism (c) [111, 112, 115].

### 3.3. 3D printed drug delivery systems

The use of 3D printing methods has been widely investigated in the biomedical field such as fabrication of customized prosthetic implants and surgical tools as well as drug delivery systems. The approaches to introduce pharmaceutical ingredients to the body are referred as drug delivery systems. Drug delivery systems have attracted lots of attention recent years, so that many works such as fluid-conveying micro/nano tubes have been carried out in this regard [116-118]. The goal is to transfer the desired amount of therapeutic agents to the planned locations with specific release rates to meet individual needs of patients [119, 120]. 3D printing of pharmaceuticals allows for flexible fabrication of multi-active dosage forms with predefined release profiles as well as an efficient production of on-demand personalized medicines [121-123]. Khaled et al. [124] performed extrusion-based printing for the production of polypills having multiple active components with a certain release profile.

FDM has been widely investigated in 3D printing of pharmaceutical components. Composites of bio-compatible filaments mainly containing PLA or PVA polymers have been predominantly used as raw materials for FDM of drug-carrying units. Several studies have been devoted to investigating capabilities of FDM in the fabrication of hollow structures such as capsular devices and coating layers as well as scaffolds with tailored holes positioned at the side [125, 126]. Moreover, in some studies, approaches like wet-spinning, melt electrospinning and hot melt extrusion (HME) process have been utilized to produce drug loaded filaments as the feedstock of FDM to directly print drug loaded structures [127-130]. Goyanes et al. [131] used a multi-nozzle printer for FDM of multiple drug-loaded filaments to produce drug delivery devices with multi-active components. Two PVA filaments, one loaded with paracetamol and the other with caffeine, were printed to make a two-part device. Simultaneous and independent drug release profiles of both drug materials were observed in the drug release test.

Binder jet printing has been also performed for drug delivery systems by applying drops of binders to a powder bed consisting of biocompatible polymers or/and active ingredients. Several investigations have been performed in this area to achieve predefined microstructures and programmed drug release profiles via printing structures with bonded regions [132-134]. Pulsatory drug release profile was achieved by tailoring drug distribution at different parts of the drug implants [135] or depositing various compositions of binder materials to confine active materials in different locations [136].

Drop-on-demand (DoD) is an IJP method in which droplets of printing materials are deposited at targeted areas. Being able to operate with a variety of printing solutions with predefined properties on different substrate materials, having robust control over drop deposition dynamics, and the possibility of using multiple nozzles or reservoirs for a single printing process make DoD as a great method for printing pharmaceutical productions with multiple active components. This technique has been applied to several pharmaceutical manufacturing procedures such as drug coating of medical implants like stents and catheters as well as 3D printing of unit doses including active ingredients [137-140]. Gupta et al. [141] applied this method for printing stimuli-responsive capsules with an aqueous core containing biomolecules and polymeric shells, which are selectively ruptured via laser irradiation to achieve a programmable release of the core materials. Fig 8 illustrates different stages of this process including the final printed three-dimensional component. Fabrication of three-dimensional capsule arrays in hierarchical structures with programmable drug release profiles via sequential deposition of biomolecules in aqueous cores was also reported.

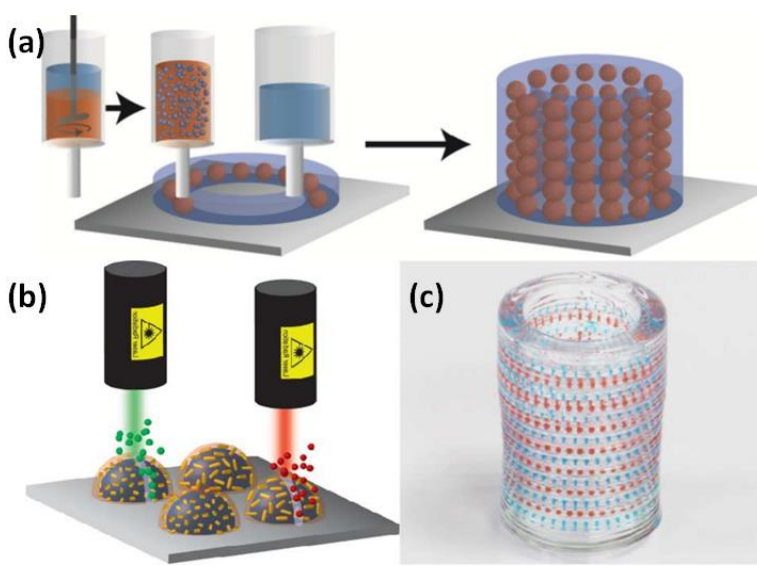

**Fig 8**. (a) Process illustration of three dimensional drug delivery system containing arrays of ink-jet printed capsule, (b) selective rupturing of polymeric shells via laser irradiation, (c) the final printed drug delivery system [141].

### 3.4. Electronics embedded 3D printed components

Additive manufacturing (AM) techniques allow for integration of other approaches like embedding of active or passive electronic components as well as conductive interconnects to fabricate three-dimensional electronic devices. The main stages of this process consist of printing dielectric structures, which can be manufactured by almost all kinds of AM methods, assembling electronic components performed by hand or pick-and-place robots, and the deposition of conductive traces, which can be done by wire embedding techniques, liquid metal paste injection, and direct writing (DW) methods such as IJP, extrusion-based methods and aerosol jet printing (AJP) [142-144].

Aerosol jet printing is a DW method which deposits an aerodynamically focused gaseous stream of aerosolized ink materials on a substrate. The properties such as adjustable deposition head direction, the ability to deposit at room temperature while maintaining a relatively high stand-off distance from the substrate, and capability of processing all kinds of materials needed for electronic devices make AJP process suitable for printing dielectric materials, semiconductors, conductive traces, and interconnectors on any 3D surface. This allows for 3D printing of components with embedded electronics, solar cells, antennae, sensors, and resistors. A thermal sintering is usually applied as a post-processing step to increase the integration of deposited particles and improve the conductivity [145-147]. Low viscosity inks are the usual raw materials for AJP. Silver nanoparticle suspensions have been used for AJP of conductive interconnectors of electronic chips in several investigations [148]. AJP of composite suspension of carbon nanotubes and silver nanoparticles has been reported for printing traces with improved conductivity [149]. A composition of aerosol printed current collecting grids of silver ink with conductive materials has been utilized for the manufacturing of optoelectronic devices [150]. 3D sub-millimeter structures with a potential to be used as passives and antennas have been manufactured by multi-material AJP of metal nanoparticles on UV curable dielectric materials, which are instantly cured after being dispensed on the substrate [151]. Fabrication of composite film of strontium titanate and alumina with AJP to make semiconducting temperature independent oxygen sensors was reported in [152]. The other advantage of multi-material AJP is the ability to print an intermediate dielectric layer between different conductive traces in the crossover areas [153].

IJP has been proved to be another approach for printing conductive interconnectors of embedded electronic devices. The use of a low-cost ink with proper conductivity, printability, less challenging post-processing demands, and compatibility with the desired substrates are the essential properties of this method [154]. Metal nanoparticles, metal-organic decompositions (MOD) and aqueous conductive solution inks have been investigated for this purpose [155-157].

To obtain the desired conductive patterns, post-printing approaches, such as thermal, photonic, or chemical sintering have been used [158]. Sequential IJP of conductive and dielectric materials has been performed on separate conductive traces in crossover areas [44, 159]. Poly-jet printing is another suitable procedure for printing equipment with embedded electronics like interactive devices [160]. Chang et al. [161] reported print-stick-peel (PSP) for the fabrication of multi-material parts of poly-jet printed components via AJP with embedded strain sensors. To avoid the pauses in the poly-jet printing process, which weakens the bonding between layers and has the adverse effect of thermal sintering of conductive ink on polymer substrates, conductive traces are printed and sintered on another substrate and then transferred to the desired location.

Stereolithography (SLA) has been also investigated in the manufacturing of functional electronics and electronics embedded parts. SLA of photopolymers containing high dielectric particles was performed to manufacture capacitors with complex geometries and certain capacitance [162]. Hybrid of SLA and DW allows for the fabrication of three-dimensional components with embedded electronics and interconnectors. Conductive traces are usually dispensed by extrusion-based printing methods. Connection pins or other conductive vias have been utilized for vertical interconnections in some applications [163-165]. Fig 9 illustrates the main stages of electronics embedded SLA hybrid production.

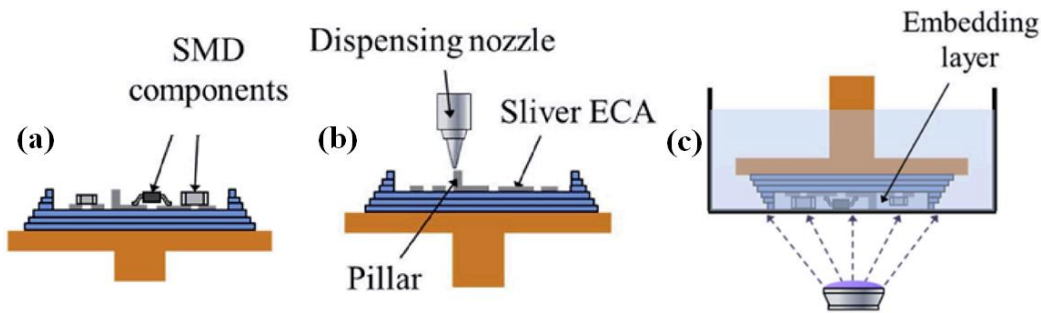

**Fig 9**. The main stages of electronics embedded SLA hybrid production [164]

FDM allows for the printing of thermoplastic materials used in the dielectric structure of three-dimensional electronic embedded parts. Aside from the proper mechanical properties of these components, they do not provide suitable resolution to embed electronic devices or conductive ink materials [166]. Furthermore, channels are needed to be formed on the surface for depositing the interconnect traces to avoid any damages to the ink caused by the deposition of subsequent layers in the process of embedding the electronics. Therefore, often subsequent subtractive methods are required to form the desired features and reach a suitable surface and dimensional quality (Fig 10.a). Thermoplastic dielectric materials also allow for the full embedding of copper wires into the surface of the electrical interconnectors by using thermal processes such as ultrasonic embedding (Fig 10.b). Bulk materials such as wires provide noticeably more electrical conductivity in comparison with the cured metallic inks. Laser micro-welding has been reported in [144] to be a suitable method to join wire interconnectors to the electronic components. In some studies, injection of low melting temperature metallic alloys into the hollow channels of structures made by FDM was used to fabricate conductive interconnectors. Hollow channels and cavities have been made by either leaving gaps during printing process [167] or by performing multi-nozzle system and printing sacrificial materials in the desired areas for the electric interconnects and removing these materials after the printing process [168] (Fig 10.c).

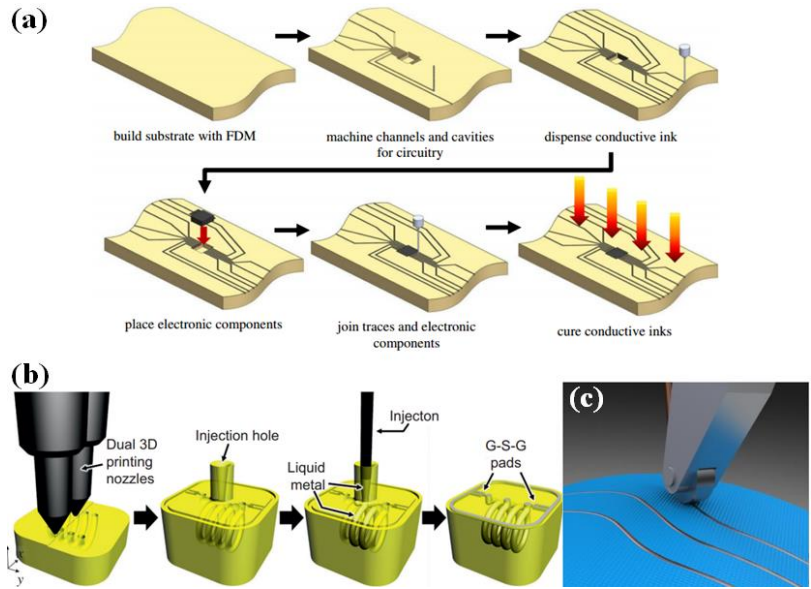

**Fig 10**. (a) Different steps of embedding electronic circuits with conductive ink interconnectors in a FDM dielectric substrate [144], (b) multi-nozzle FDM followed by melted metal injection for conductive parts [168], (c) ultrasonic wire embedding to thermoplastic substrates [169].

Binder jet printing has also been utilized for the dielectric structure of electronic embedded parts with the combination of AJP for printing conductive interconnecting traces. Selectively binding of powder material was performed to shape the three-dimensional structure of the component. Additionally, an exhausting system was applied to remove the powder in unbounded regions to create cavity and channels in order to place electrical parts and dispense interconnector materials [170]. SLS of polymer materials was also integrated with DW methods, including AJP and extrusion-based method of dispensing silver inks, to create conductive traces embedded components [171].

UC allows for fully embedding electronic parts into a solid metallic structure in a relatively low-temperature process. Machining processes can be utilized to create pockets to embed electronic devices as well as interconnectors [172]. The use of DW of silver ink was reported in [173] for printing conductive traces as an antenna for UC electronic embedded parts. The possibility of creating electronic embedded dense metal matrix components was investigated by screen printing of dielectric materials, which have the potential of embedding electronic components into metallic substrates and directly encasing them with UC of metallic foils in a solid part [174, 175]. Fig 11 illustrates these two different kinds of embedding electronics in this process.

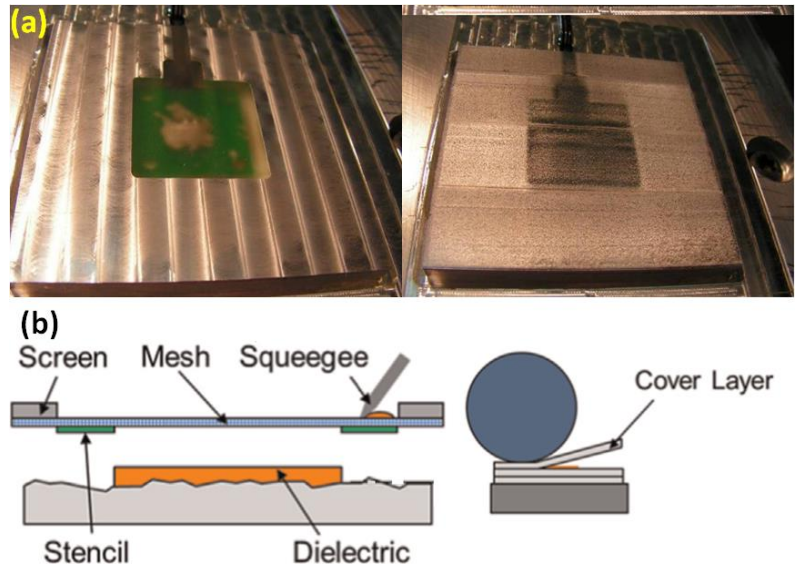

**Fig 11**. (a) Hybrid of UC and pocketing for embedding electronic devices [172], (b) direct UC embedding of electronic materials in metal matrices [175].

Electromechanical systems are the integration of electronics and mechanical parts which either use the electronic signals to trigger movements or generate electronic signals actuated by moving parts. AM of electromechanical systems has the capability of rapid fabrication of structures with complicated and customized shapes and dimensions. A micro-scale bistable vibration energy harvester (VEH) was designed and fabricated by several deposition processes followed by some patterning steps [176]. Aguilera et al. [169] developed a hybrid fabrication method to produce a 3-phase DC motor by five different stages including assembling ball bearings and electrical components, fused deposition of thermoplastic material to subsequently build and encapsulate different parts of the device, wire embedding, and laser micro welding of wires and electrical components for the interconnections (Fig 12.a). Fuller et al. [177] used two and three dimensional IJP of metal nanoparticle colloids to fabricate microelectromechanical systems (MEMS) such as electrothermal actuators and planar cantilevered structures (Fig 12.b).

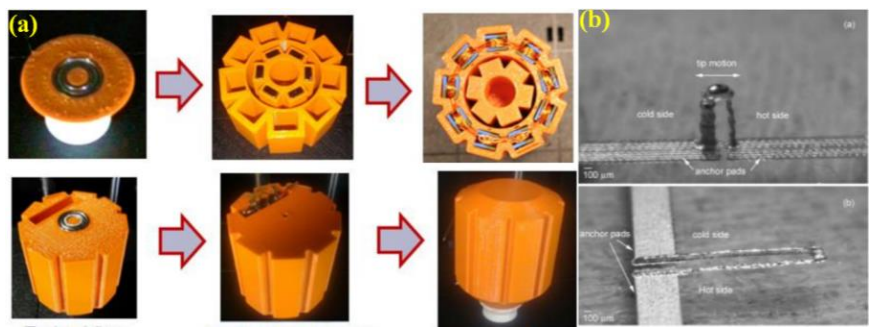

**Fig 12**. (a) Different stages of hybrid AM of a 3-phase DC motor [169], (b) ink-jet printed actuators with vertical and horizontal morphologies [177].

## 4. Summary and Conclusions

Different experimental processes including modification of AM and their combinations with other manufacturing processes targeting more efficient production of multi-material and composite products have been reviewed in this study. Several investigated potentials of AM in being combined with other manufacturing processes have been mentioned. Moreover, various functions of these products in different industries including medical devices, electronics, biomedical implementations, and robotics have been summarized. Another purpose of this literature review was to investigate how and in what orders different manufacturing operations can be combined to reach a component with unconventional structures and practicality. For this purpose, AM was considered as the main focus of this study and other additional procedures were mentioned as post-processes, pre-processes or concurrent ones with AM. The immense effects of AM methods in the mentioned areas with their simplicity and flexibility have been observed for the past two decades. However, many potentials of AM methods in manufacturing and design development are still to be investigated.